\documentclass[manuscript]{aastex6}

\newcommand{\fpp}[2]{\frac{\partial #1}{\partial #2}}
\newcommand{\vctr}[1]{\mbox{\boldmath $#1$}}
\newcommand{\mrm}[1]{\mathrm{#1}}

\begin{document}


\title{Reconnection--Condensation Model for Solar Prominence Formation}


\author{Takafumi Kaneko\altaffilmark{1} and Takaaki Yokoyama\altaffilmark{2}}
\altaffiltext{1}{Institute for Space-Earth Environmental Research, Nagoya University,
  Furo-cho, Chikusa-ku, Nagoya, Aichi 464-8601, Japan; kaneko@isee.nagoya-u.ac.jp}
\altaffiltext{2}{Department of Earth and Planetary Science,The University of Tokyo, 7-3-1 Hongo, Bunkyo-ku, Tokyo 113-0033, Japan}

\begin{abstract}

We propose a reconnection--condensation model
in which topological change in a coronal magnetic field 
via reconnection triggers radiative condensation, thereby resulting in prominence formation.
Previous observational studies have suggested 
that reconnection at a polarity inversion line of a coronal arcade field
creates a flux rope that can sustain a prominence;
however, they did not explain the origin of cool dense plasmas of 
prominences.
Using three-dimensional magnetohydrodynamic simulations
including anisotropic nonlinear thermal conduction 
and optically thin radiative cooling, 
we demonstrate that reconnection 
can lead not only to flux rope formation but also 
to radiative condensation under a certain condition.
In our model, this condition is described by the Field length,
which is defined as the scale length 
for thermal balance between radiative cooling and thermal conduction.
This critical condition depends weakly on 
the artificial background heating.
The extreme ultraviolet emissions synthesized with our simulation results 
have good agreement with observational signatures reported
in previous studies. 

\end{abstract}


\section{Introduction} \label{sec:intro}

Solar prominences, also known as filaments, are cool dense plasma clouds 
that are typically 100 times cooler and
denser than the ambient hot corona.
They are one of the basic structures in the corona, 
and are strongly related to solar eruptions, magnetohydrodynamic
(MHD) waves, and coronal heating.
Despite the importance of prominences in solar physics, the mechanism of their formation, 
i.e., the origin of magnetic structures and cool dense plasmas,
is not fully understood.
Prominences always appear along polarity inversion lines (PILs) 
across which magnetic polarity at the photosphere is reversed
\citep{Martin1998SoPh,Mackay2010SSRv}.
Previous observational and theoretical studies
proposed that a flux rope sustaining a prominence is formed via
magnetic reconnection along a PIL
\citep{vanBallegooijen1989ApJ,Gaizauskas1997ApJ,MartensZwaan2001ApJ,Yang2016ApJ}.
However, this reconnection scenario does not explain 
the origin of cool dense plasmas alone.

Radiative condensation is one mechanism that show promise in explaining  
how cool dense plasmas are generated in the hot tenuous corona.
The observations by the 
{\it Solar Dynamics Observatory} Atmospheric Imaging Assembly (SDO/AIA)
revealed a temporal and spatial shift in the peak intensities
of multiwavelength extreme ultraviolet (EUV) emissions toward 
low temperatures during prominence formation events.
This could be the evidence for radiative condensation \citep{Berger2012ApJ,Liu2012ApJ}.

From a theoretical viewpoint,
the corona is thermally unstable even against linear perturbations of sufficiently long wavelength. On the other hand, nonlinear triggers are necessary
to grow the thermal instability until the establishment of prominence-corona
transition by overcoming the strong stabilizing effect of thermal conduction.
Several models involving radiative condensation via nonlinear triggers
have been proposed.
One such model is the evaporation-condensation model
\citep{Mok1990ApJ,Antiochos1991ApJ,
Antiochos1999ApJ,Karpen2001ApJ,Karpen2003ApJ,Karpen2005ApJ,
Karpen2006ApJ,KarpenAntiochos2008ApJ,Xia2011ApJ,Luna2012ApJ,
XiaChenKeppens2012ApJ,KeppensXia2014ApJ,XiaKeppens2016ApJ}. 
In this model, 
radiative condensation is triggered by an enhancement in the
plasma density via an evaporation process driven by strong steady artificial heating 
at a footpoint of a coronal loop.
The crucial factor in this model is  
steady footpoint heating to drive chromospheric evaporation.
An estimation in \citet{Ashwanden2000ApJ} using EUV data suggested
the presence of nonuniform coronal heating localized at the
footpoint of coronal loops. However,
hot evaporated flows from the chromosphere have not been detected in
prominence formation events.

Other radiative condensation models have adopted reconnection
via the footpoint motions of coronal arcade fields
as an alternative trigger to evaporation via footpoint heating.
A numerical study by \citet{KanekoYokoyama2015ApJ} proposed
a reconnection--condensation model
with in situ radiative condensation
triggered by magnetic reconnection. It does not involve any additional 
artificial footpoint heating or chromospheric evaporation.
In this model, a flux rope comprising closed magnetic loops 
is formed via reconnection. The interior of the closed loops
does not gain heat from the exterior via thermal conduction. 
After suffering a cooling-dominant thermal imbalance,
the interior of the closed loops cools continuously until it
reaches the temperature of the prominence.
This model was demonstrated in \citet{KanekoYokoyama2015ApJ} 
using 2.5-dimensional MHD simulations 
including optically thin radiative cooling 
and thermal conduction along magnetic fields.
In \citet{Linker2001JGR} and \citet{Zhao2017ApJ}, prominence formation
via direct injection of chromospheric cool plasmas after reconnection 
was modeled in two-dimensional simulations.
These models are not consistent with the in situ condensations 
observed recently \citep{Berger2012ApJ,Liu2012ApJ}.
However, it is possible to explain the levitation of 
chromospheric plasmas with a rising helical flux rope
\citep{Okamoto2008ApJ,Okamoto2009ApJ,Okamoto2010ApJ}.

In this paper, we investigate the reconnection--condensation model
using three-dimensional MHD simulations.
In the previous simulations performed by \cite{KanekoYokoyama2015ApJ}, thermal conduction
along the axial magnetic fields of a flux rope, which are orthogonal to the simulation
domain in their settings, was neglected due to 2.5-dimensional assumption,
i.e., temperature gradient perpendicular to the simulation domain was assumed to be zero.
In the three-dimensional situation, relaxation via 
thermal conduction might be more efficient,
and there might be a condition whereby radiative condensation 
is limited by this effect.
To check the validity of the reconnection--condensation model 
and to obtain such a critical condition 
for prominence formation,
we performed three-dimensional MHD simulations
including nonlinear anisotropic thermal conduction,
optically thin radiative cooling and gravity. 

The numerical settings are described in Section \ref{sec:setting}. 
In Section \ref{sec:res}, the results of the simulations and the images of 
the EUV emissions synthesized through the filters of SDO/AIA are shown.
We discuss the results in Section \ref{sec:discussion},
and a conclusion of the paper is given in Section \ref{sec:conclusion}.

\section{Numerical Settings} \label{sec:setting}

The simulation domain is a rectangular box 
whose Cartesian coordinates $(x,y,z)$  
extend to $-12$ $\mathrm{Mm} < x < 12$ $\mathrm{Mm}$, $0 < y < 40$ $\mathrm{Mm}$,
and $0 < z < 65$ $\mathrm{Mm}$, where the $y$-direction 
corresponds to the height and the $xz$-plane is parallel to the horizontal plane.

The initial corona is under hydrostatic stratification
with a uniform temperature ($T_\mathrm{cor}=1~\mathrm{MK}$)
and a uniform gravity ($g_\mathrm{cor}=270~\mathrm{m/s^{2}}$).
The initial density profile is given as
\begin{equation}
  n=n_{\mathrm{cor}}\exp \left[ -\frac{y}{L_{s}}\right],
\end{equation}
where $n$ is number density,
$n_{\mathrm{cor}}=10^{9}~\mathrm{cm^{-3}}$ is the number density at $y=0$ and
$L_{s}=k_{B}T_{\mathrm{cor}}/(mg_{\mathrm{cor}})=30~\mathrm{Mm}$
is the coronal scale height.
The initial magnetic field is a linear force-free arcade and is given as
\begin{eqnarray}
  B_{x} &=& -\left( \frac{2L_{a}}{\pi a}\right)B_{a} 
  \cos \left( \frac{\pi x}{2L_{a}} \right) \exp \left[ -\frac{y}{a} \right], \\
  B_{y} &=& B_{a}\sin \left( \frac{\pi x}{2L_{a}} \right) \exp \left[ -\frac{y}{a} \right], \\
  B_{z} &=& -\sqrt{1-\left( \frac{2L_{a}}{\pi a}\right)^{2}}  
  B_{a}\cos \left( \frac{\pi x}{2L_{a}} \right) \exp \left[ -\frac{y}{a} \right],
\end{eqnarray}
where $B_{a}=6~\mathrm{G}$, $L_{a}=12~\mathrm{Mm}$, and $a=30~\mathrm{Mm}$
(see Fig. \ref{res1} (a)). The PIL is located at $x=0$.
The three-dimensional MHD equations including nonlinear anisotropic 
thermal conduction and optically thin radiative cooling are solved numerically
as follows: 
\begin{equation}
  \fpp{\rho }{t}+\nabla \cdot \left(\rho \vctr{v}\right)=0,
  \label{eq_mass_nc}
\end{equation}
\begin{equation}
  \fpp{\left(\rho \vctr{v}\right)}{t}+\nabla \cdot \left( \rho \vctr{v}\vctr{v}
  +p\vctr{I}-\frac{\vctr{B}\vctr{B}}{4\pi }+\frac{B^2}{8\pi
  }\vctr{I} \right) -\rho \vctr{g}=0,
  \label{eq_momentum_nc}
\end{equation}
\begin{eqnarray}
  \frac{\partial }{\partial t}\left( e_\mrm{th}+\frac{1}{2}\rho
  \vctr{v}^{2} +\frac{B^2}{8\pi }
  \right)
  +\nabla \cdot \left[ \left( e_\mrm{th} + p +\frac{1}{2}\rho
    \vctr{v}^{2}\right)\vctr{v}+\frac{c}{4\pi }\vctr{E} \times \vctr{B}
    \right] \nonumber \\
  =\rho \vctr{g}\cdot \vctr{v}
  +\nabla \cdot \left(\kappa T^{5/2}\vctr{b}\vctr{b}\cdot \nabla T \right)
  -n^{2}\Lambda (T)+H,
  \label{eq_energy_nc}
\end{eqnarray}
\begin{equation}
  e_\mrm{th}=\frac{p}{\gamma -1},
  \label{eth}
\end{equation}
\begin{equation}
  T=\frac{m}{k_{B}}\frac{p}{\rho },
  \label{te_def}
\end{equation}
\begin{equation}
  \fpp{\vctr{B} }{t}=-c\nabla \times \vctr{E},
  \label{eq_induction}
\end{equation}
\begin{equation}
  \vctr{E}=-\frac{1}{c}\vctr{v}\times \vctr{B}
  +\frac{4\pi \eta }{c^{2}} \vctr{J},
  \label{eq_ohm}
\end{equation}
\begin{equation}
  \vctr{J}=\frac{c}{4\pi }\nabla \times \vctr{B},
  \label{eq_current}
\end{equation}
where $\mrm{\kappa =2\times 10^{-6}~erg~cm^{-1}~s^{-1}~K^{-7/2}}$
is the coefficient of thermal conduction, $\vctr{b}$
is a unit vector along the magnetic field,
$\Lambda (T)$ is the radiative loss function of
optically thin plasma, $H$ is the background heating rate,
and $\eta $ is the magnetic diffusion rate.
We use the same radiative loss function as that used
in \citet{KanekoYokoyama2015ApJ}.
However, there is no temperature cutoff, rather, we assume a   
dependence of $T^{3}$ below $10^{4}~\mathrm{K}$. 
To achieve the initial thermal equilibrium,
the background coronal heating proportional to the 
magnetic energy density is set to $H=\alpha _{A}B^{2} $,
where $B$ is the magnetic field strength and
$\alpha _{A}=1.5 \times 10^{-6}~\mathrm{s^{-1}}$ is a constant coefficient.
Note that $\alpha _{A}$ can be a constant because the coronal scale height $L_{s}$
is the same as the magnetic decay length $a$ such that
\begin{eqnarray}
  &~& n^{2}\Lambda (T_\mathrm{cor})=H \\
  \Leftrightarrow &~& n_{\mathrm{cor}}^{2}\Lambda (T_{\mathrm{cor}}) \exp \left[ \frac{2y}{L_{s}} \right]
  =\alpha _{A} B_{a}^{2}\exp \left[\frac{2y}{a}\right] \\
  \Leftrightarrow &~&\alpha _{A}
  =\frac{n_{\mathrm{cor}}^2 \Lambda(T_\mathrm{cor})}{B_{a}^2}.
\end{eqnarray}
For fast magnetic reconnection,
we adopt the following form of the anomalous resistivity
\citep[e.g.][]{YokoyamaShibata1994ApJ}:
\begin{eqnarray}
  \eta & = & 0, ~~~\left(J < J_{c}\right)\\
  \eta & = & \eta _{0}\left(J/J_{c}-1\right)^{2}, ~~~\left(J \ge J_{c}\right)
\end{eqnarray}
where $\eta _{0}=\mrm{3.6\times 10^{13}~cm^{2}~s^{-1}}$ and
$J_{c}=\mrm{25~erg^{1/2}~cm^{-3/2}~s^{-1}}$.
We restrict $\eta $ to $\eta _{\mrm{max}}=\mrm{1.8\times 10^{14}~cm^{2}~s^{-1}}$.
To drive reconnection at the PIL of the arcade field, 
footpoint velocities perpendicular and parallel to the PIL are given as 
\begin{eqnarray}
  v_{x} &=& -v_{0}(t) \sin \left( \frac{\pi x}{2L_{a}}\right) 
  \exp \left[-\left( \frac{z}{L_{a}} \right) ^{2} \right], \label{eq_vx}\\
  v_{y} &=& 0, \label{eq_vy}\\
  v_{z} &=& v_{x}, \label{eq_vz}
\end{eqnarray}
where $t$ represents time, and $v_{0}(t)$, the speed dependent on time, 
is set in the region below $y=0$.
The speed $v_{0}(t)$ is given as 
\begin{eqnarray}
  v_{0}(t) &=& v_{00}, ~~(0<t<t_{1}) \\
  v_{0}(t) &=& v_{00}\frac{t_{2}-t}{t_{2}-t_{1}}, ~~(t_{1} \le t<t_{2}) \\
  v_{0}(t) &=& 0,~~(t \ge t_{2})   \label{fvel3}
\end{eqnarray}
where $v_{00}=6~\mathrm{km/s}$, $t_{1}=4320~\mathrm{s}$, and
$t_{2}=4560~\mathrm{s}$.
In the numerical study by \citet{KanekoYokoyama2015ApJ}, a parameter survey
on the direction of parallel motion was performed using 2.5-dimensional simulations.
They concluded that radiative condensation is more likely to occur
in the case of anti-shearing motion
(in which the direction of parallel motion is opposite to that of the magnetic shear)
rather than in the case of shearing motion
(in which the direction of parallel motion is the same as that of the magnetic shear).
In the present study, we investigate the cases of anti-shearing motion
using three-dimensional simulations.
The magnetic fields below $y=0$ are computed by the induction equation
with the footpoint velocities given as Eqs. (\ref{eq_vx}) - (\ref{fvel3}).
The free boundary condition is 
applied to the magnetic fields at the lower boundary. 
The gas pressure and density below $y=0$ are assumed 
to be unchanged
in hydrostatic equilibrium at a uniform temperature of $1~\mathrm{MK}$.
The free boundary condition is applied to all the variables 
at the upper boundary.
The anti-symmetric boundary condition is applied 
to $v_{x},v_{z},B_{x},$ and $B_{z}$, 
and the symmetric boundary condition is applied to the other variables 
at the boundaries in the $x$-direction.  
Because the system evolves in two-fold rotational symmetry around the $y$-axis at
$(x,z)=0$, we set symmetry boundary condition for a rotation of 180 degrees
in the $xy$-plain at $z=0$ to reduce numerical amount.  
We also assume the same rotational symmetry at $(x,z)=(0,65~\mathrm{Mm})$.
In the following figures, plots in the range of $-65~\mathrm{Mm}<z<65~\mathrm{Mm}$
are shown.

The numerical scheme is a four-stage Runge--Kutta method \citep{Vogler2005AA} and a fourth-order central finite difference method with artificial viscosity \citep{Rempel2014ApJ}. Thermal conduction is explicitly solved using a super-time-stepping method with second-order temporal and spatial accuracy \citep{Meyer2012MNRAS,Meyer2014}. The grid spacing size is $120$ $\mathrm{km}$ everywhere.

\section{Results} \label{sec:res}
\begin{figure}[htbp]
  \begin{center}
    \begin{tabular}{c}
      \begin{minipage}{1.0\hsize}
        \begin{center}
          \includegraphics[bb=0 0 465 270,scale=0.9]{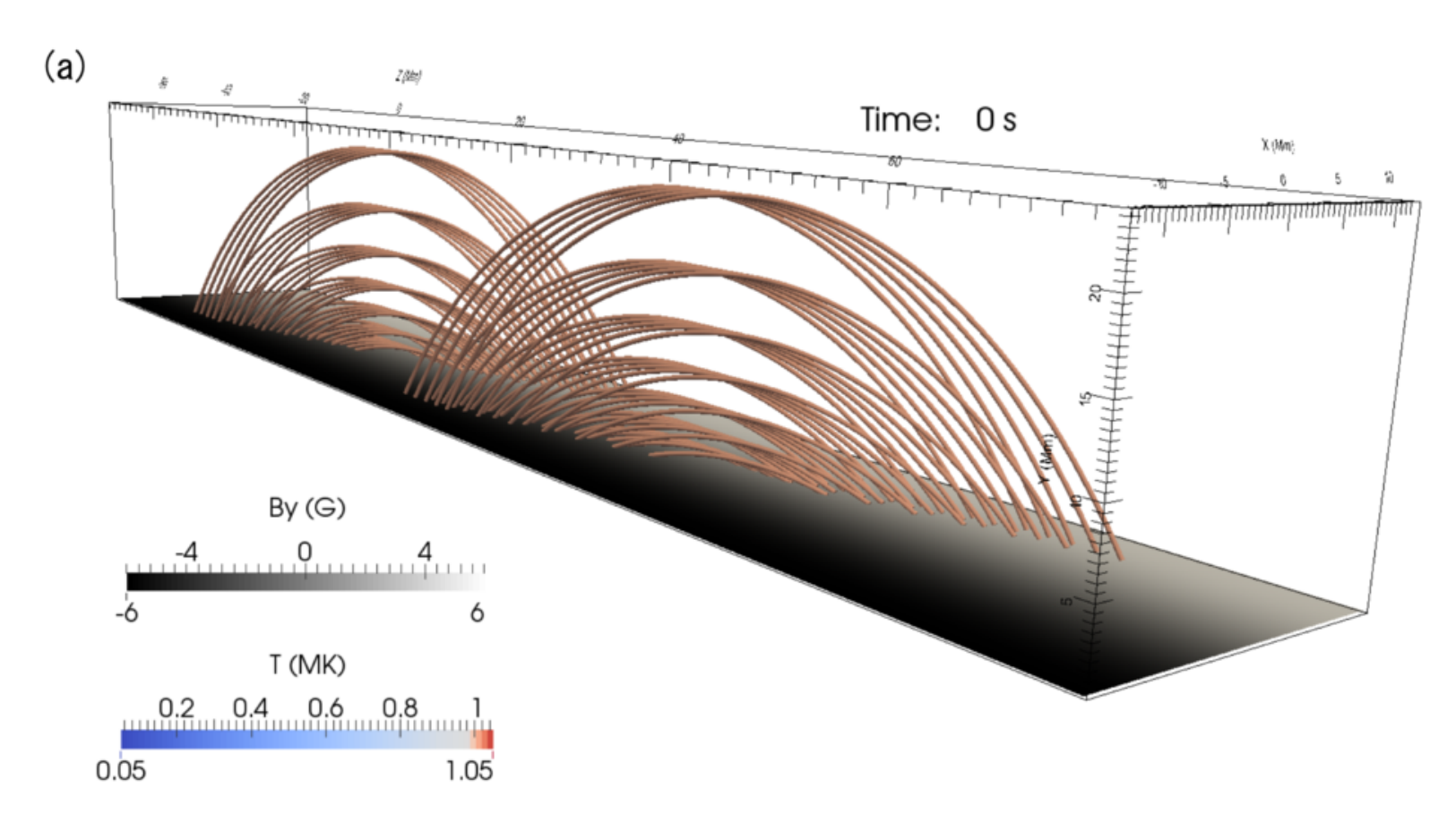}
        \end{center}
      \end{minipage} \\      
      \begin{minipage}{1.0\hsize}
        \begin{center}
          \includegraphics[bb=0 0 465 270,scale=0.9]{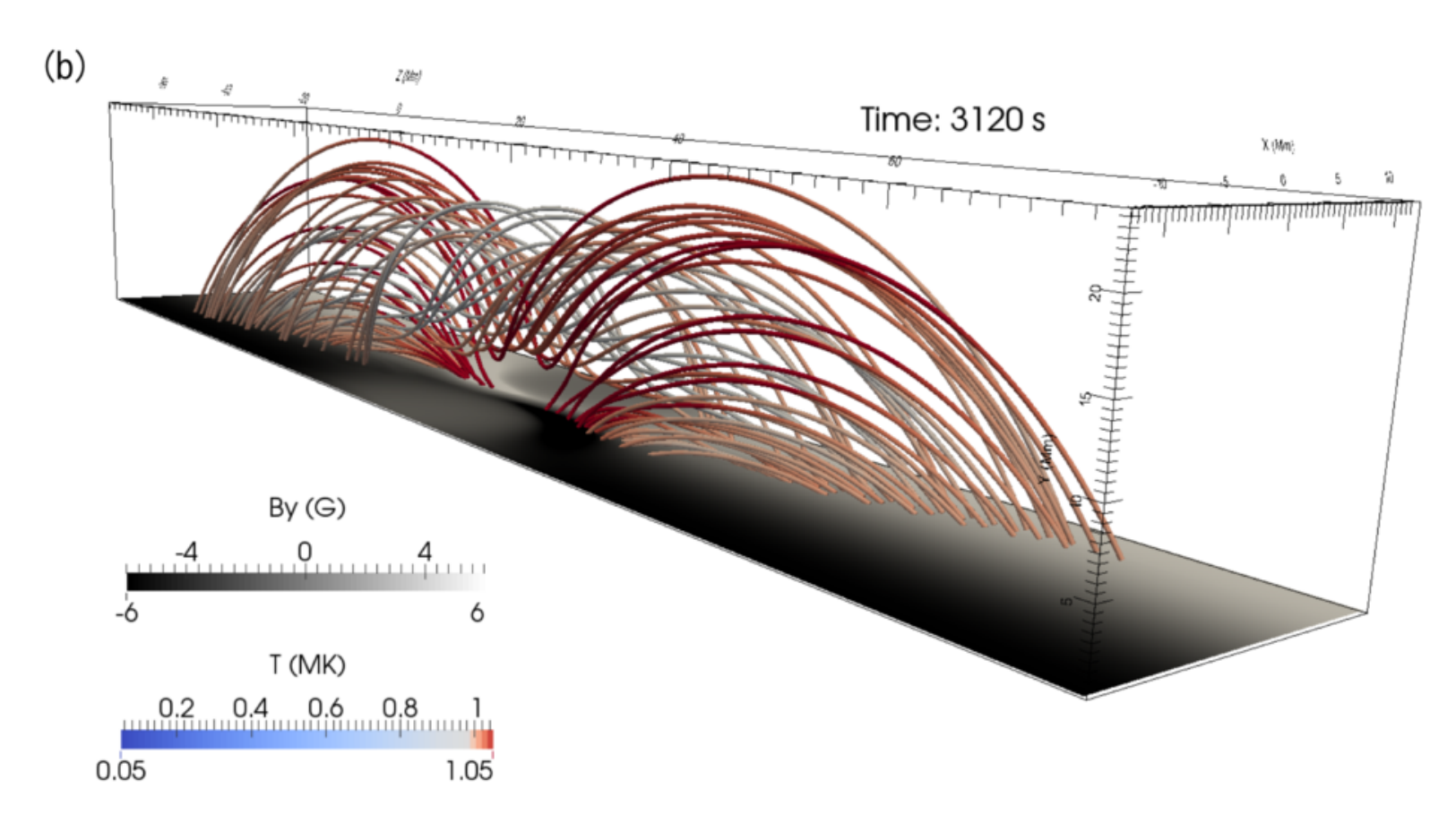}
        \end{center}
      \end{minipage}
    \end{tabular}
    \caption{Simulation results. Panel (a) shows the 
      initial condition. Panel (b) shows the snapshot
      after the formation of a flux rope.
      The lines indicate the magnetic field and their color 
      represents the temperature. 
      The grayscale at the bottom boundary represents the signed strength of 
      the magnetic field perpendicular to the surface.}
    \label{res1}
  \end{center}
\end{figure}

\begin{figure}
  \begin{center}
    \begin{tabular}{c}
      \begin{minipage}{1.0\hsize}
        \begin{center}
          \includegraphics[bb=0 0 465 270,scale=0.9]{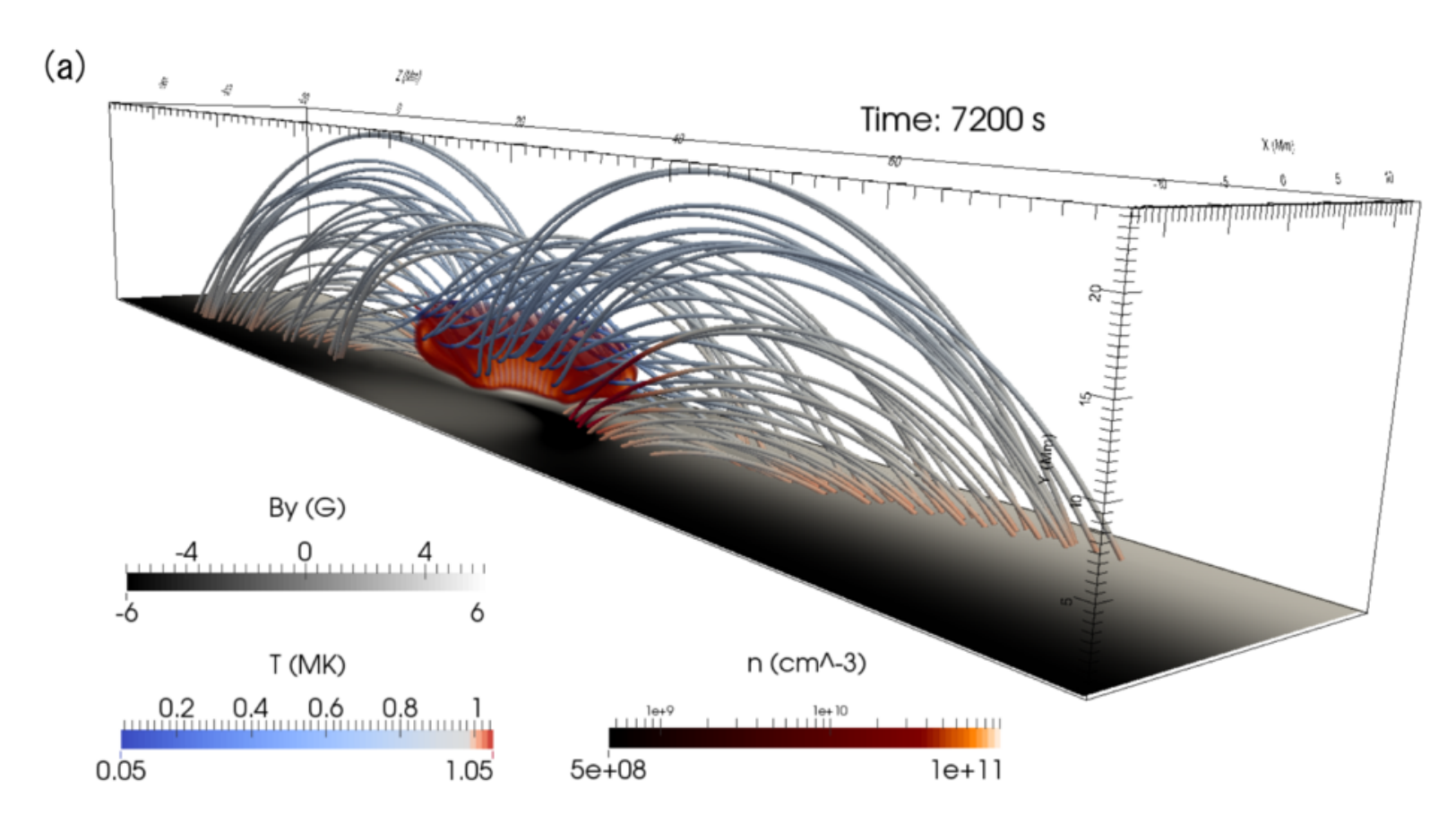}
        \end{center}
      \end{minipage} \\
      \begin{minipage}{1.0\hsize}
        \begin{center}
          \includegraphics[bb=0 0 465 270,scale=0.9]{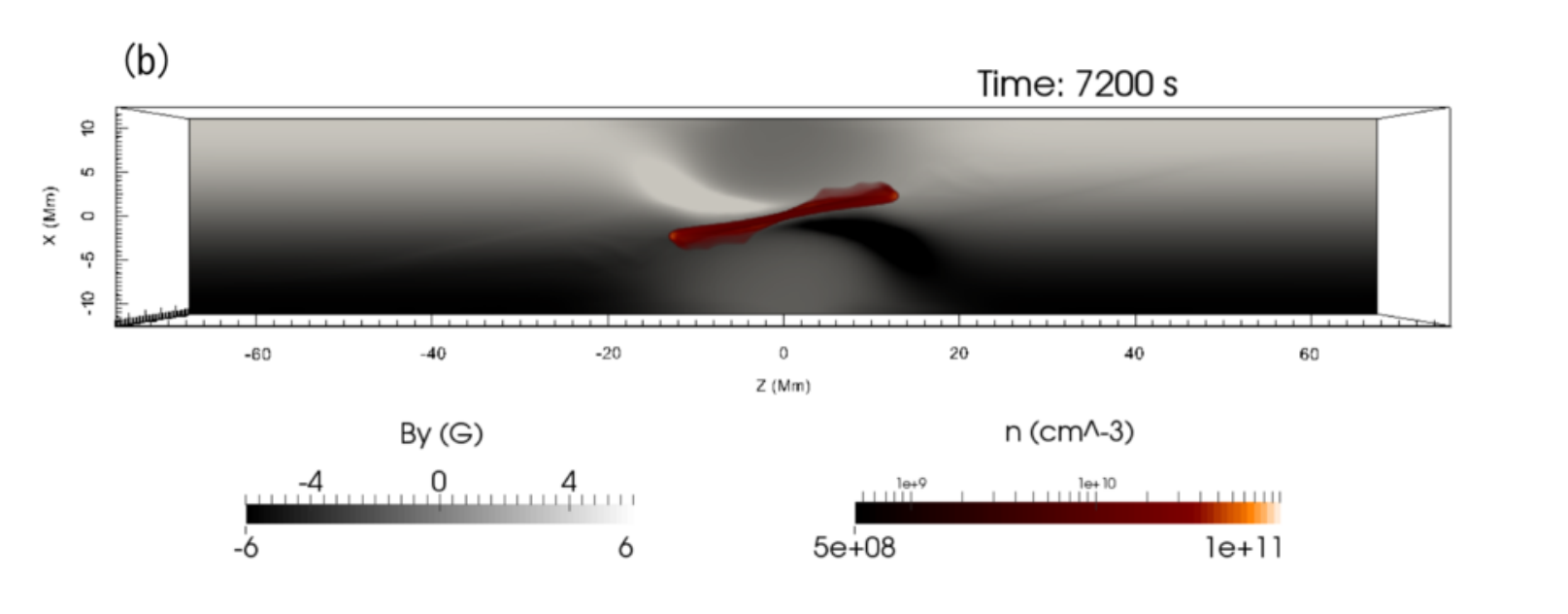}
        \end{center}
      \end{minipage}
    \end{tabular}
    \caption{Simulation results after radiative condensation.
      Panel (a) shows a side view with the magnetic field lines,
      and Panel (b) shows the top view without magnetic field lines.
      The area with a high density ($n > 4.0 \times 10^{9}~\mathrm{cm^{-3}}$) 
      is shown in red volume rendering. The lines, colors, and grayscale 
      represent the same as those shown in Fig. \ref{res1}.}
    \label{res2}
  \end{center}
\end{figure}

Figures \ref{res1} and \ref{res2} are snapshots of our simulation result.
The initial arcade field (Fig. \ref{res1} (a)) evolves
into a flux rope structure owing to reconnection via the footpoint motion 
(Fig. \ref{res1} (b)).
The flux rope traps dense plasmas at the lower corona,
and the radiative cooling inside the flux rope overwhelms 
the background heating. 
Owing to reconnection, the length of the magnetic loops becomes longer, 
and the relaxation effect of thermal conduction 
along the long magnetic loops becomes weaker.
The enhanced radiative loss is not fully compensated for
by the thermal conduction along the long reconnected magnetic loops, 
leading to radiative condensation (Fig. \ref{res2} (a)). 
The condensed plasmas are accumulated into the magnetic dips of 
the flux rope because of gravity. 
Owing to the location of the dips, the dense plasmas concentrate
above the PIL to form a filament structure as shown
in the top view of Fig. \ref{res2} (b).

\begin{figure}
  \begin{center}
    \includegraphics[bb=0 0 287 197,scale=1.0]{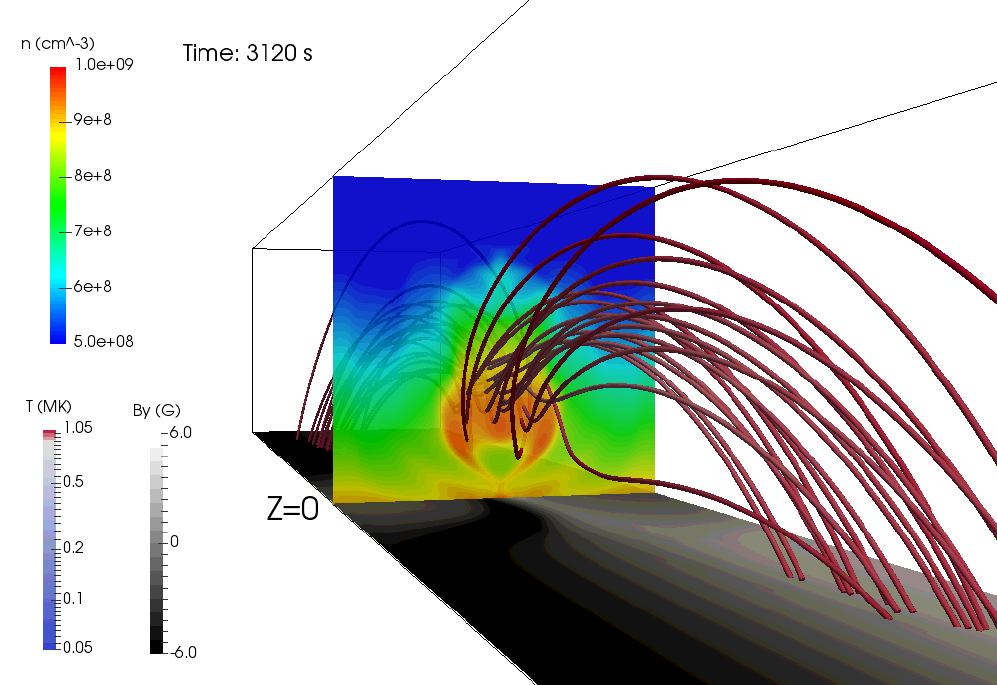}
    \caption{Density distribution inside the flux rope at $t = 3120~\mathrm{s}$.
      Colors on the $xy$-plane at $z=0$ represent density.
      The lines and grayscale at the bottom boundary represent the same 
      as those in Fig. \ref{res1}.}
    \label{den0065}
  \end{center}
\end{figure}

\begin{figure}
  \begin{center}
    \includegraphics[bb=0 0 340 226,scale=1.0]{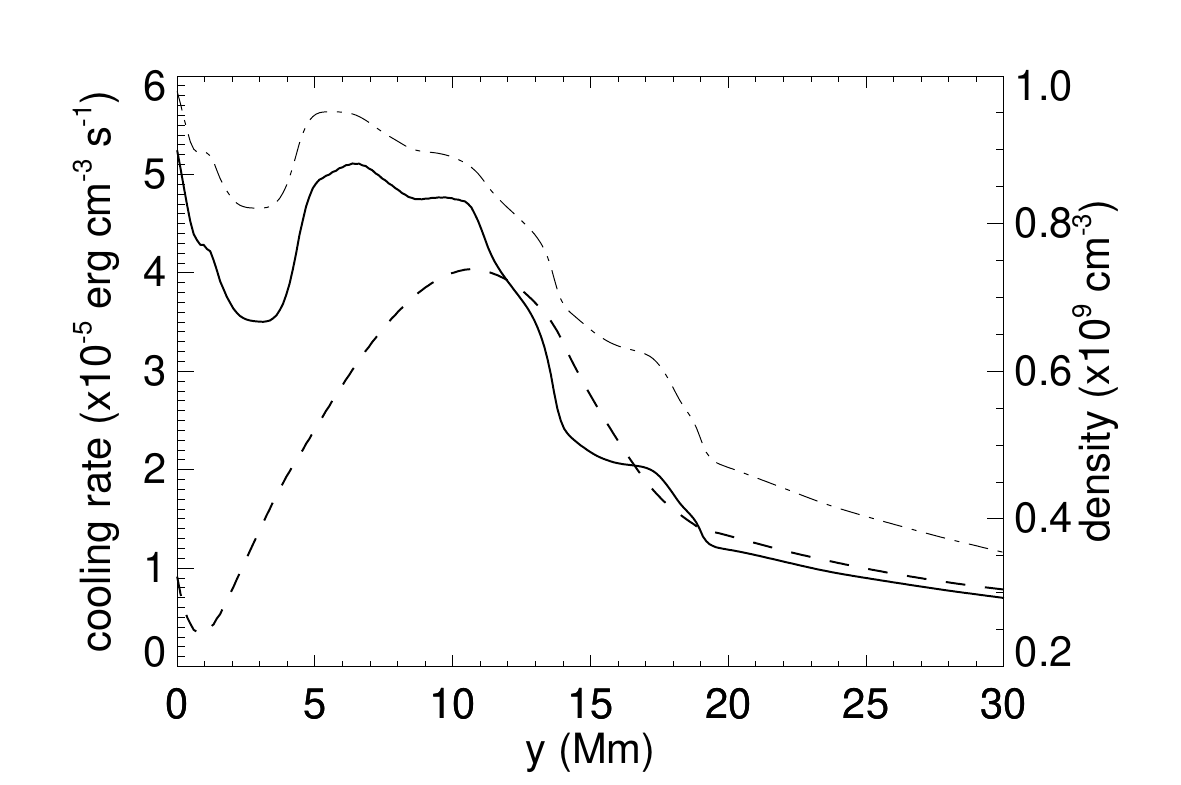}
    \caption{Cooling and heating rate along the $y$-axis in the $xy$-plane at $z=0$
      and $t=3120~\mathrm{s}$.
      Solid and dashed lines represent the cooling and heating rates, respectively.
      The dash-dotted line represents the density profile.}
    \label{clht0065}
  \end{center}
\end{figure}

Figure \ref{den0065} shows the density distribution at $t = 3120~\mathrm{s}$ in
  the $xy$-plain at $z=0$ with magnetic field lines of the flux rope.
  The dense plasmas initially at the lower altitude are trapped: they are 
  subsequently lifted up by the ascending flux rope. Figure \ref{clht0065}
  shows the profiles of cooling rate, heating rate, and density along $y$-axis
  in the $xy$-plane.
  The cooling rate is enhanced by
  the dense plasmas trapped inside the flux rope ($y=3-13~\mathrm{Mm}$), and
  it overwhelms the background heating rate.
  The enhanced radiative loss inside the flux rope is the source of radiative condensation.

\begin{figure}[htbp]
  \begin{center}
    \includegraphics[bb=0 0 340 425,scale=0.8]{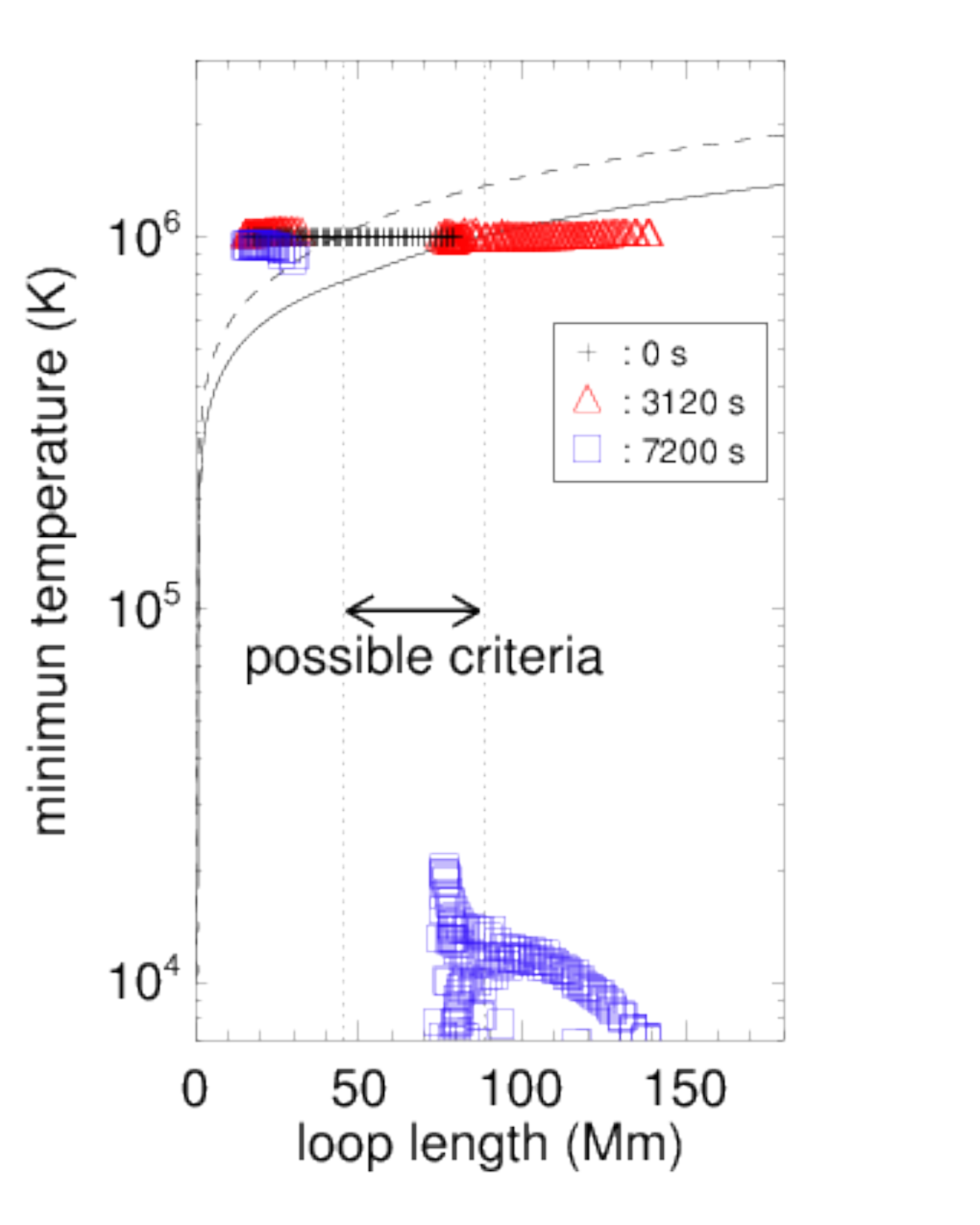}
    \caption{Relationship between the loop length 
      and the minimum temperature of individual magnetic loops 
      at different times. 
      Solid and dashed lines represent
      $\lambda _{F}(T,n_{t})$ and $\lambda _{F}(T,n_{b})$ respectively.
      Black crosses, red triangles and blue squares represent the state 
      at the different times, as shown in the inset.}
    \label{len_te}
  \end{center}
\end{figure}

The relationship between the length and the minimum temperature
is shown in Fig. \ref{len_te} for each individual magnetic field line.
At the initial state, the temperature is uniform at $10^{6}~\mathrm{MK}$ 
(black crosses in Fig. \ref{len_te}).
The loop length becomes roughly double after reconnection 
($t=3120$ s, red triangles), and the longer loops 
suffer from radiative condensation ($t=7200$ s, blue squares).
The critical length for radiative condensation 
can be explained by the Field length \citep{Field1965ApJ}, which given as
\begin{equation}
  \lambda _{F} (T,n) \approx \sqrt{\frac{\kappa T^{7/2}}{n^{2}\Lambda (T)}},
  \label{field}
\end{equation}
and the critical condition is
  \begin{equation}
    L>\lambda _{F},
  \end{equation}
  where $L$ is the length of the magnetic loops.
The solid and dashed lines in Fig. \ref{len_te} 
represent the Field lengths $\lambda _{F}(T,n_{t})$ and 
$\lambda _{F}(T,n_{b})$, 
where $n_{t}=5.0\times 10^{8}~\mathrm{cm^{-3}}$ 
and $n_{b}=1.0\times 10^{9}~\mathrm{cm^{-3}}$ are 
the densities around the top of the tallest arcades subject to 
reconnection, and at the bottom boundary, respectively.
The possible criteria in Fig. \ref{len_te} are estimated based on
maximum thermal conduction effect (of $1~\mathrm{MK}$) versus
possible variation of radiative cooling associated with the density profile
along the field lines.
Our simulation results show that
the Field length can approximately 
explain the condition of radiative condensation.

\begin{figure}[htbp]
  \begin{tabular}{ccc}
    \begin{minipage}{0.3\hsize}
      \includegraphics[bb=0 0 283 425,scale=0.6]{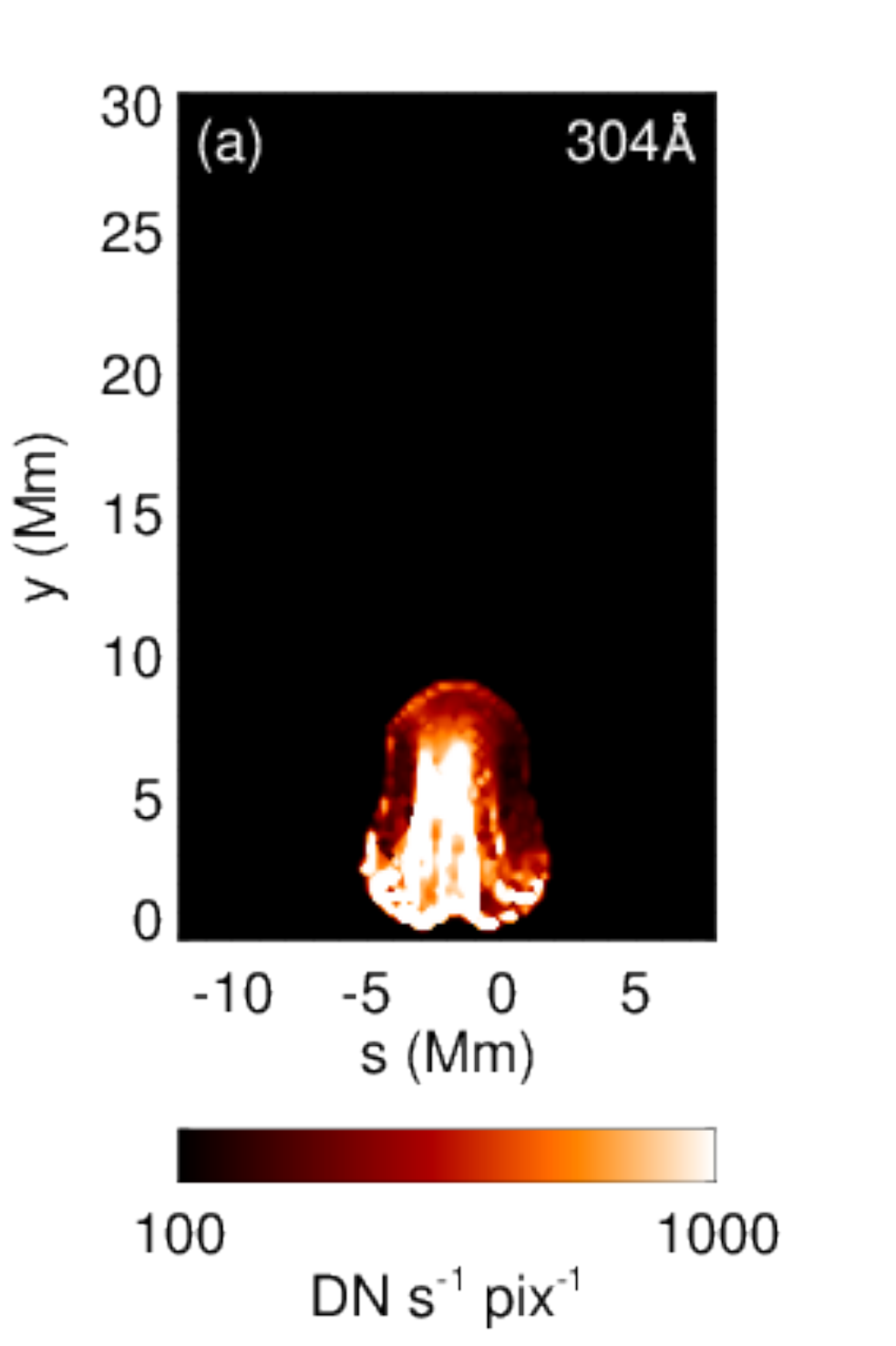}
    \end{minipage}
    \begin{minipage}{0.3\hsize}
      \includegraphics[bb=0 0 283 425,scale=0.6]{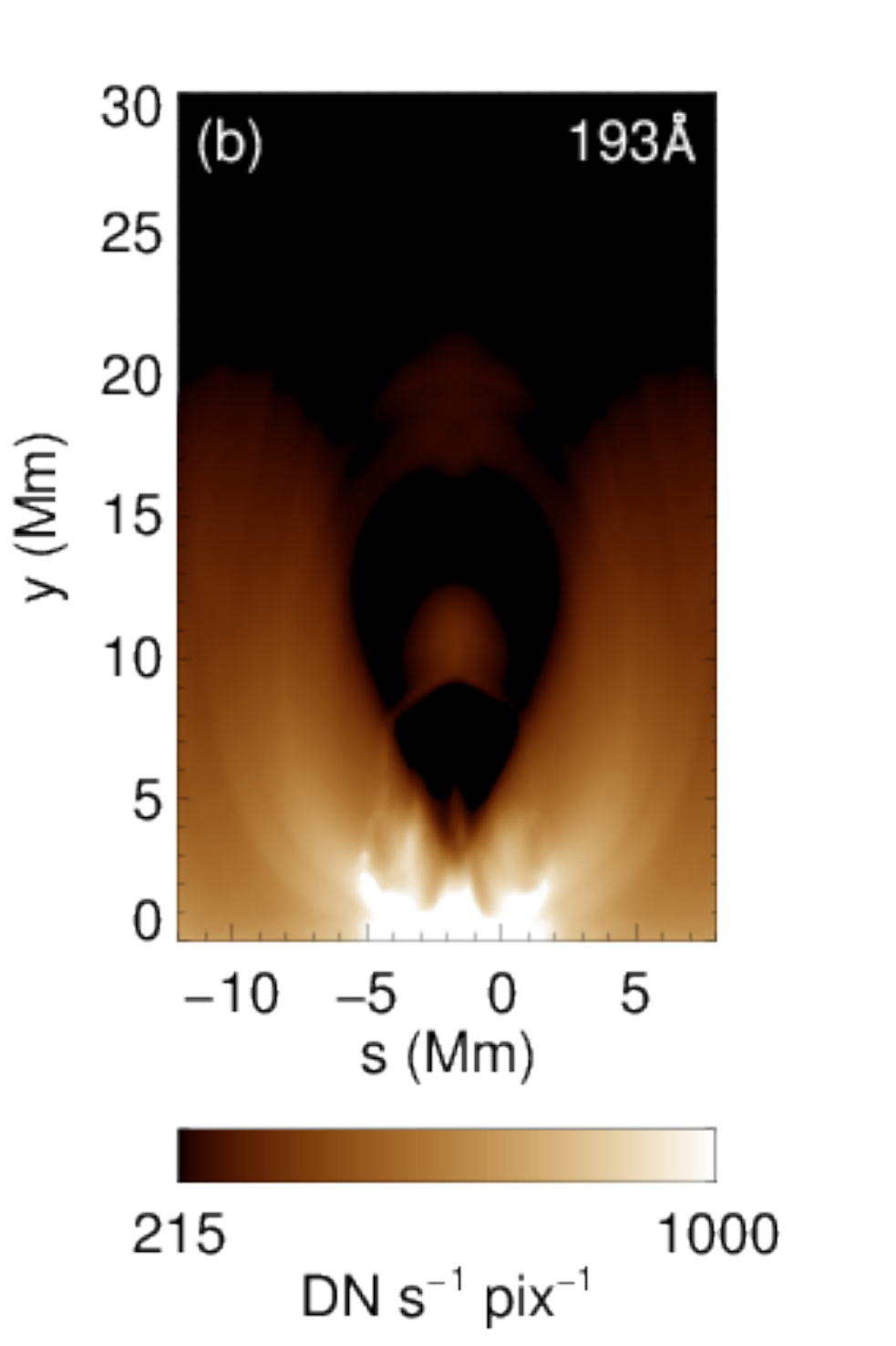}
    \end{minipage}
    \begin{minipage}{0.3\hsize}
      \includegraphics[bb=0 0 283 425,scale=0.6]{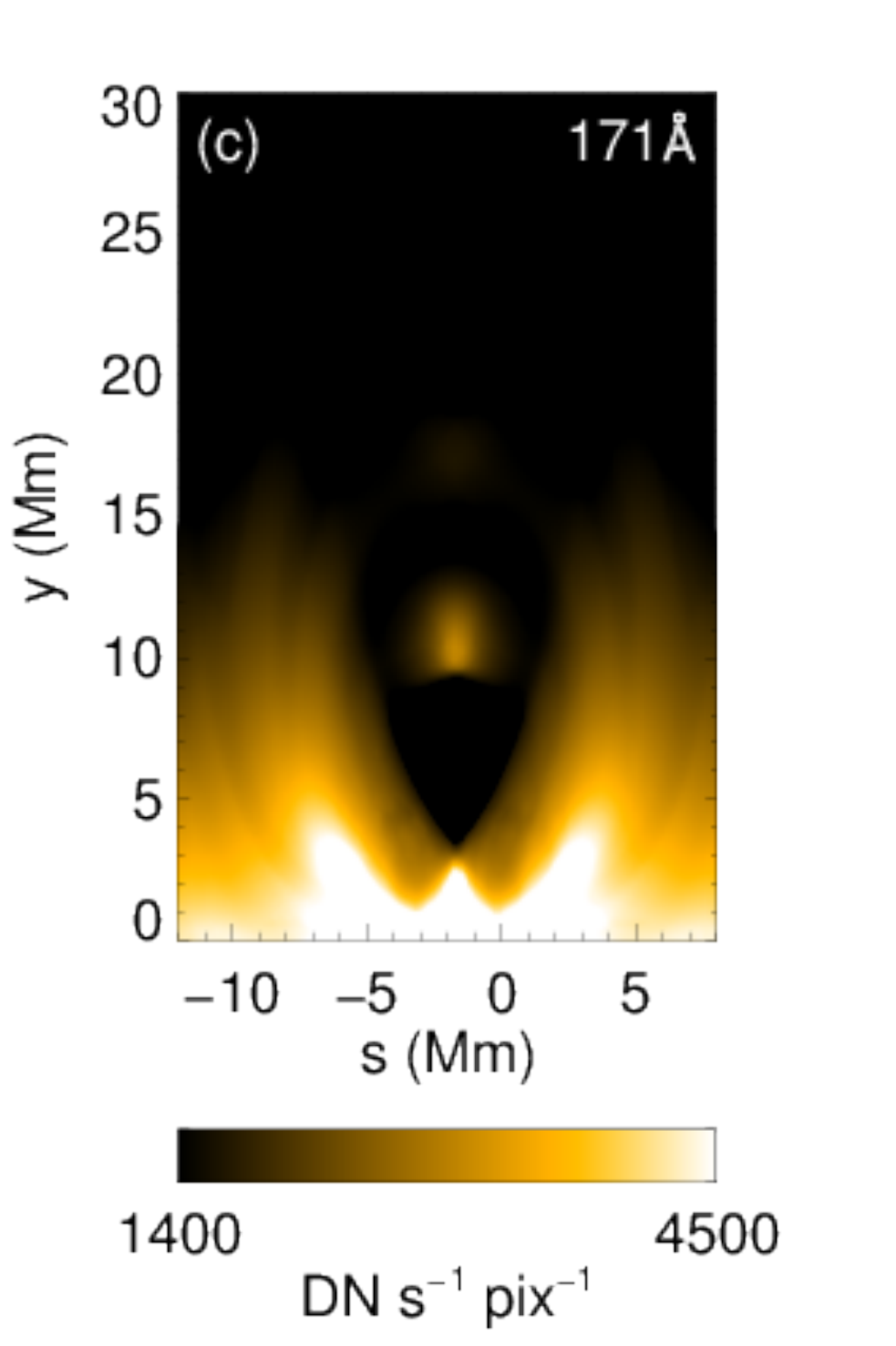}
    \end{minipage}    
  \end{tabular}
  \caption{EUV emission synthesized through the SDO/AIA filters 
    at $t = 7200~\mathrm{s}$. Panels (a), (b) and (c) represent 
    the images obtained at $304 \mathrm{\AA}$, $193 \mathrm{\AA}$, 
    and $131 \mathrm{\AA}$, respectively.
    The angle between the line of sight and the $z$-axis is chosen to be 
    $5^{\circ }$ for the synthesis. The coordinate $s$ is inclined at $5^{\circ }$
    to the $x$-axis.}
  \label{syn}
\end{figure}

\begin{figure}[htbp]
  \begin{center}
    \includegraphics[bb=0 0 720 479,scale=0.6]{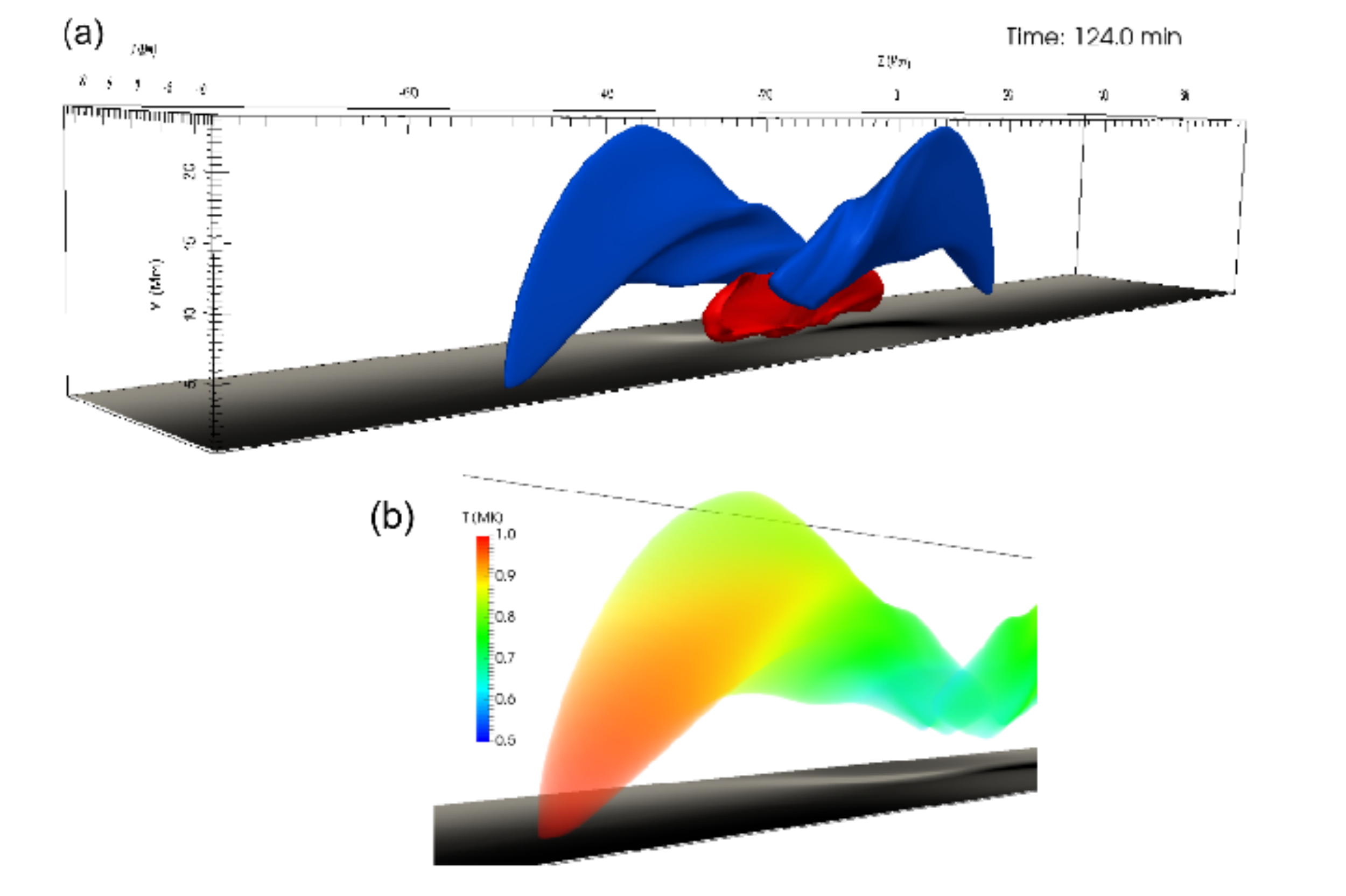}
    \caption{Panel (a) : three-dimensional morphology of the cavity and prominence. Blue and red contours represent density surface of $n=3.5\times 10^{8}~\mathrm{cm^{-3}}$ and $n=1.0\times 10^{10}~\mathrm{cm^{-3}}$, respectively. Panel (b) : distribution of temperature inside the cavity shown by volume rendering.}
    \label{mor}
  \end{center}
\end{figure}

\begin{figure}[htbp]
  \begin{center}
    \includegraphics[bb=0 0 425 425]{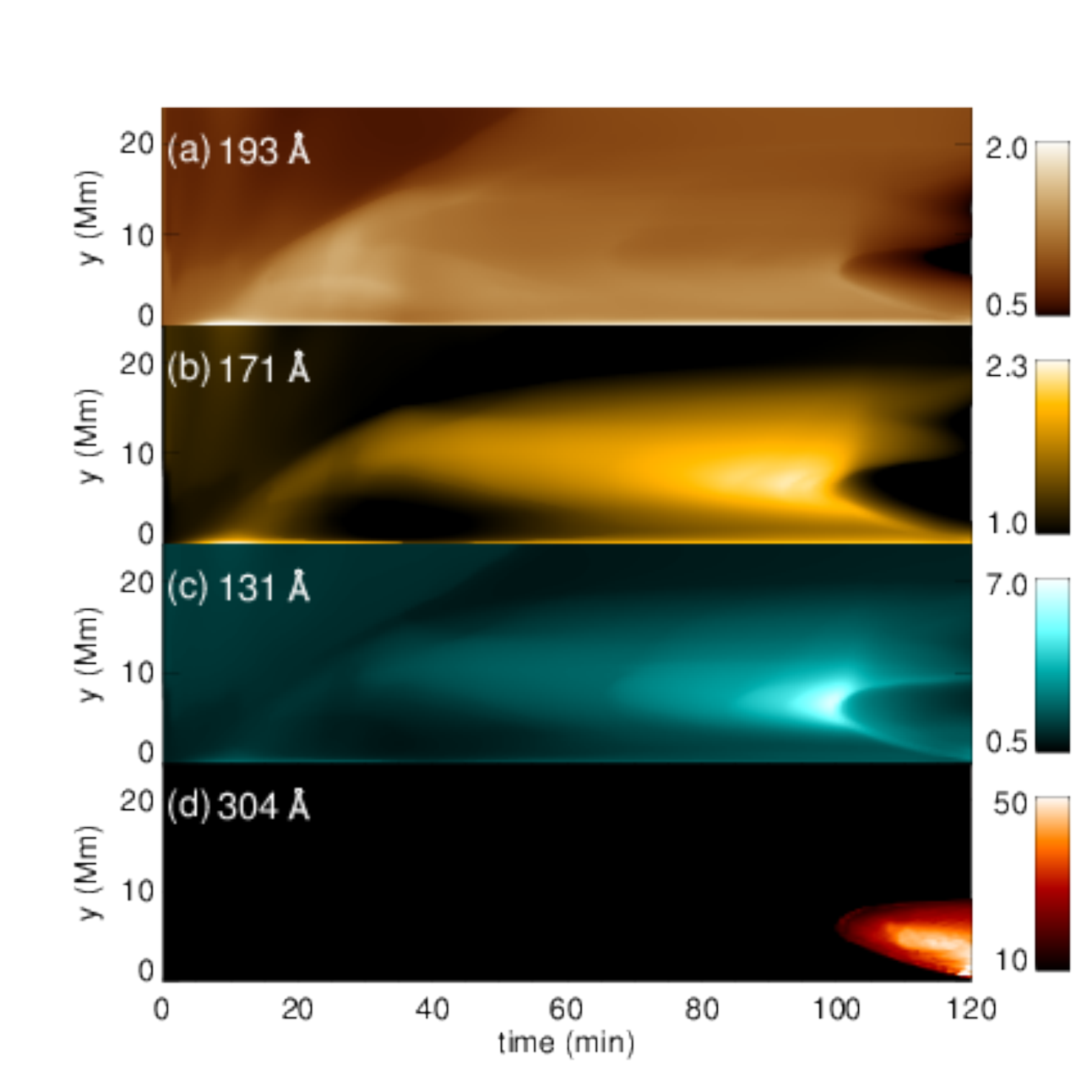}
    \caption{Time--height plot of the synthesized EUV emission ratios 
      through SDO/AIA filters. The emission in the region
      from $s=-5~\mathrm{Mm}$ to $2~\mathrm{Mm}$
      in Fig. \ref{syn} is averaged. The emission at time = 0 s 
      is the reference of the ratio for each filter. 
      Panels (a), (b), (c), and (d) represent the images 
      of $193~\mathrm{\AA}$, 
      $171~\mathrm{\AA}$, $131~\mathrm{\AA}$, and $304~\mathrm{\AA}$,
      respectively.}
    \label{syn_tevol}
  \end{center}
\end{figure}

Figure \ref{syn} shows the EUV emissions synthesized through the filters 
of SDO/AIA.
A dark cavity surrounding the prominence is formed
in the emissions of the coronal temperatures at 
193 and 171 $\mathrm{\AA}$.
A low density cavity is formed 
as a natural consequence of mass conservation
because all the prominence mass in our model comes from the surrounding corona.
In Fig. \ref{mor} (a), the regions surrounded by blue surfaces 
represent the three-dimensional morphology of the cavity
in our simulation. We find that the shape of the cavity resembles the
wing of a seagull: this is because
the cavity is formed along the sheared coronal magnetic field lines.
As shown in Fig. \ref{mor} (b), temperature inside the cavity ranges from $0.6$ to $1.0~\mathrm{MK}$. The variation of temperature has little influence on the shape of the cavity in 193 $\mathrm{\AA }$, because
the temperature response function is almost constant in this range of temperature.
The variation of temperature can affect the shape of the cavity in
171 $\mathrm{\AA }$ where the response function has a peak around $0.8~\mathrm{MK}$
and is not constant.
However, the shape of cavity in $171~\mathrm{\AA }$ is
quite similar to that in $193~\mathrm{\AA }$ (see Fig. \ref{syn} (b) and (c)).
It is concluded that the depletion of mass mostly contributes to
the formation of the cavity.
The maximum density and the minimum temperature of our simulated prominence are $10^{11}~\mathrm{cm^{-3}}$ and $0.06~\mathrm{MK}$, which are $100$ times higher and lower
than those in the initial corona, respectively.
The averaged density and the volume in the region where temperature
is lower than $0.1~\mathrm{MK}$ are $2 \times 10^{10}~\mathrm{cm^{-3}}$ 
and $8 \times 10^{26}~\mathrm{cm^{3}}$, respectively.
Although our model does not include the chromosphere, the resultant mass
reaches the observed lower limit 
of typical prominence densities and volumes \citep{Labrosse2010SSRv}.
This is because the volume of the cavity is much larger than that of the prominence,
as shown in Fig. \ref{mor}.
Figure \ref{syn_tevol} shows the time evolution of the EUV emissions.
During radiative condensation,
the intensity peak shows a temporal shift from 171 $\mathrm{\AA}$
(coronal temperature) to 304 $\mathrm{\AA}$ (prominence temperature),
which is qualitatively consistent with the observations made
by \citet{Berger2012ApJ}.
The time scale until condensation starts in our simulation
is rather closer to that in active regions
\citep[several tens of minutes,][]{Yang2016ApJ}
than that in quiet regions \citep[several hours,][]{Berger2012ApJ,Liu2012ApJ}.
This is because the initial density in our simulations
($5-10 \times 10^{8}~\mathrm{cm^{-3}}$)
is higher than the typical coronal density in quiet regions
($10^{8}~\mathrm{cm^{-3}}$).
Note that the 211 $\mathrm{\AA}$ filter image is not synthesized here
because the initial coronal temperature is $1~\mathrm{MK}$ in our simulation.
A 131 $\mathrm{\AA}$ filtergram 
is used as the signature of the transition temperature
between the corona and the prominence.
The 193 $\mathrm{\AA}$ emission 
is bright during the reconnection phase ($t= 10$ - $20~\mathrm{min}$), 
which reflects 
the temperature increase from $1~\mathrm{MK}$ to $1.2~\mathrm{MK}$ 
inside the flux rope via reconnection heating
and the density increase via the 
levitation of the lower coronal plasma. 

\section{Discussion} \label{sec:discussion}

Our study propose a new possible process to meet the condition
for thermal instability in the magnetized plasma 
compared to the previous models, e.g., evaporation--condensation model.
In our reconnection--condensation model, 
the length of the magnetic loops exceeds the Field length 
as a result of magnetic reconnection.
The Field length in our model depends on the density and temperature,
which are nearly unchanged from those in the initial coronal state. 
On the other hand,
in the evaporation--condensation model, 
the loop length does not change and the Field length becomes shorter
owing to mass supply from the evaporated flows \citep{Xia2011ApJ}.
These two different routes to condensation can both occur in actual solar atmospheres.
We speculate that our reconnection--condensation model is more appropriate to explain
prominence formation in quiet regions where the strong evaporation can not be expected.

Future observational studies should
investigate the length of reconnected loops
in terms of their extension beyond the Field length.
The Field length is longer in quiet regions 
and shorter in active regions
owing to differences in the typical densities.
This may explain
the difference between typical length of 
quiescent prominences ($\sim 100~\mathrm{Mm}$) 
and active region prominences ($\sim 10~\mathrm{Mm}$).

\begin{figure}
  \begin{center}
    \includegraphics[bb=0 0 610 259,scale=0.7]{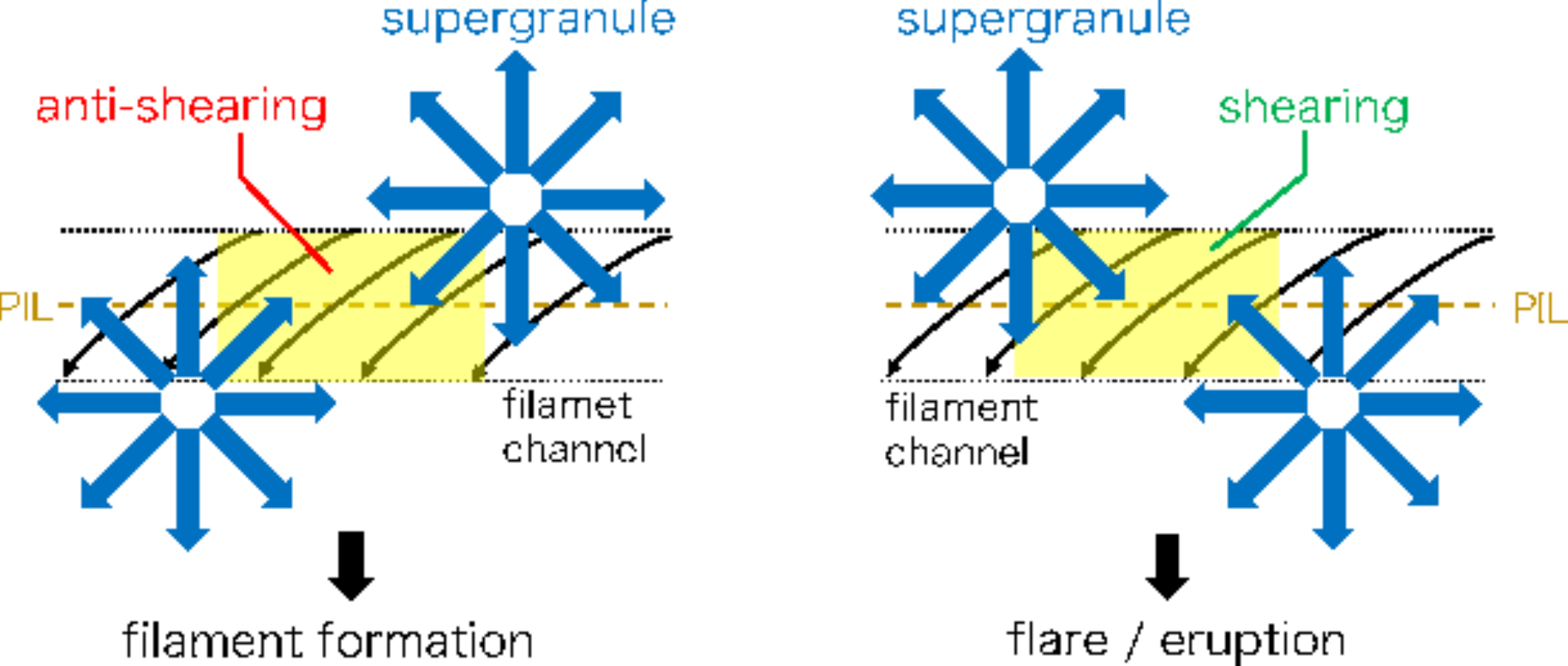}
    \caption{Possible relationship between the relative position of supergranules and
      the direction of footpoint motion. The top view of filament channels is shown,
      and black curved arrows represent magnetic fields. Blue arrows represent
      diverging flows of the supergranules. Yellow regions suffer from converging motion
      and anti-shearing or shearing motion depending on relative position
      of the supergranules.}
    \label{superg}
  \end{center}
\end{figure}

A possible origin of the converging and anti-shearing motions is the
interaction of diverging
flows of supergranules crossing a PIL \citep{Rondi2007AA,Schmieder2014AA},
which have a mean speed of $0.3~\mathrm{km/s}$.
Because the typical lifetime of supergranules is one day,
the migration distance of a magnetic element is approximately  
$20~\mathrm{Mm}$.
In this study, we use a footpoint motion with a speed of
$6~\mathrm{km/s}$ for $4000~\mathrm{s}$.
Because the migration distance
is consistent with the observational values, the amount of
reconnected fluxes in our simulations is reasonable.
We speculate that both anti-shearing and shearing motions are created
depending on the relative position of supergranules against the direction of the magnetic
shear of the filament channel, as shown in Fig. \ref{superg}. The results of a parameter
survey conducted in \citet{KanekoYokoyama2015ApJ} show that anti-shearing
results in radiative condensation, whereas shearing causes eruptions.
Future observational studies on filament formation
should investigate the distribution of supergranules around filament channels.

\begin{figure}[htbp]
  \begin{center}
    \includegraphics[bb=0 0 340 425,scale=0.8]{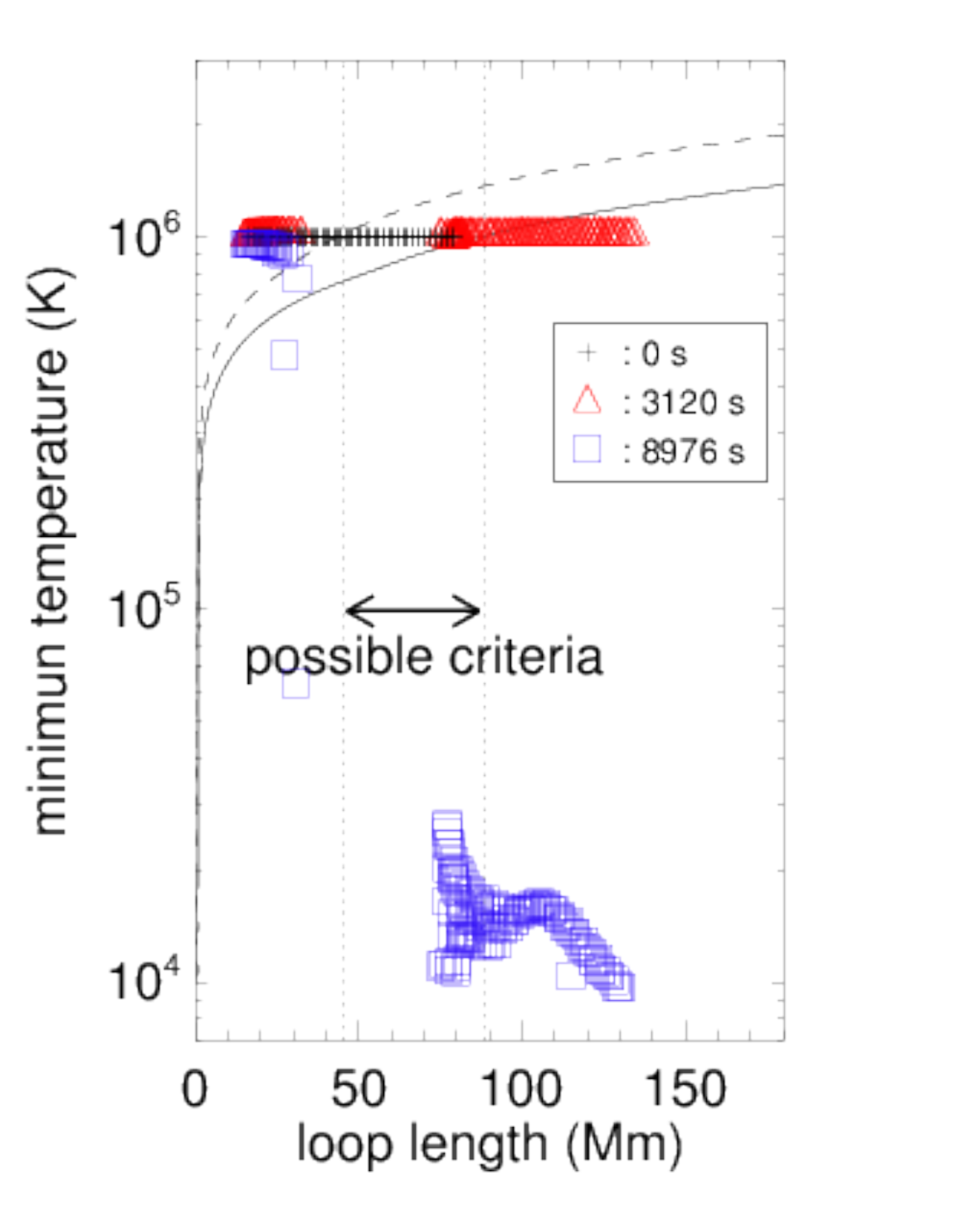}
    \caption{Relationship between the loop length 
      and the minimum temperature of individual magnetic loops 
      in the case of heating model $H=\alpha _{B}n B$. 
      Black crosses, red triangles and blue squares represent the state 
      at the different times, as shown in the inset.
      Solid and dashed lines represent the same as those shown in Fig. \ref{len_te}.}
    \label{len_te_B}
  \end{center}
\end{figure}

To check the dependence on the artificial background heating,
we performed an additional simulation using a different type of background heating,
$H=\alpha _{B}n B$, where
$\alpha _{B}=9.2 \times 10^{-15}~\mathrm{erg^{-1/2}~cm^{-3/2}~s^{-1}}$ is
set for the initial thermal balance.  
Figure \ref{len_te_B} shows relationship between the loop length and
the minimum temperature of magnetic loops.
The critical condition of the Field length is valid even in this heating model.
By comparing the results of two different heating models, we find
a common tendency that the longer loops reach the lower temperatures
(see blue squares in Figs. \ref{len_te} and \ref{len_te_B}).
This is because the stabilizing effect of thermal conduction is weaker
in the longer loops. The deviation from this tendency in the shorter loops can be
a multi-dimensional effect, where, as the prominence descends by gravity,
the shorter loops below the prominence start to condense due to compression.
A difference between the results of two heating models is the temperature
of prominences.
The prominence temperature in the heating model of $H=\alpha _{B} n B$ is
higher because the heating rate inside the prominence increases
as the density increases.
In this study, we examined only two heating models.
In future, other types of heating models proposed by
previous studies \citep{Mandrini2000ApJ} need to be investigated.
Simulation of condensation using a 
self-consistent coronal heating model 
is also a candidate for future studies.

As mentioned in Section \ref{sec:res}, the mass of our simulated prominence 
satisfies only the lower limit of the typical prominence mass. 
To obtain more mass, one possible approach is to construct a
longer flux rope using multiple reconnections.
Because several converging points are observed around PILs
owing to the interactions between supergranules
 \citep{Schmieder2014AA},
multiple reconnection events are plausible.
Another possibility is an additional mass supply from chromospheric jets 
\citep{Chae2003ApJ} or a siphon-like mechanism driven by
a strong pressure gradient during condensation
\citep{PolandMariska1986SoPh,ChoeLee1992SoPh,Karpen2001ApJ,Xia2011ApJ}.

Understanding the mechanism of mass circulation 
between the corona, prominences and the chromosphere is an important
issue when discussing the mass budget of prominences 
and coronal mass ejections (CMEs).
This issue may also be related to the mechanisms of 
recurrent prominence formation and homologous CMEs.
A numerical simulation by \citet{XiaKeppens2016ApJ} reproduced
  a prominence with fragmented interior structures
and indicated that a continuous mass supply via 
evaporation can sustain the mass cycle between the chromosphere 
and the corona via prominences, however, such a steady strong 
evaporation has not yet been observed.
In our study, the simulated prominence did not show interior fine structures
reported in the previous studies
\citep{Berger2008ApJ,Hillier2012ApJa,KeppensXia2015ApJ}.
Perturbations to trigger the Rayleigh-Taylor instability might be necessary.
In addition, our model excludes the chromosphere, and the transfer of mass occurs  
only from the corona to the prominence. 
To discuss the mass transport system,
our model must include the chromosphere in the future.

\begin{figure}
  \begin{center}
    \begin{tabular}{cc}
      \begin{minipage}{0.5\hsize}
        \begin{center}
          \includegraphics[bb=0 0 283 340,scale=0.85]{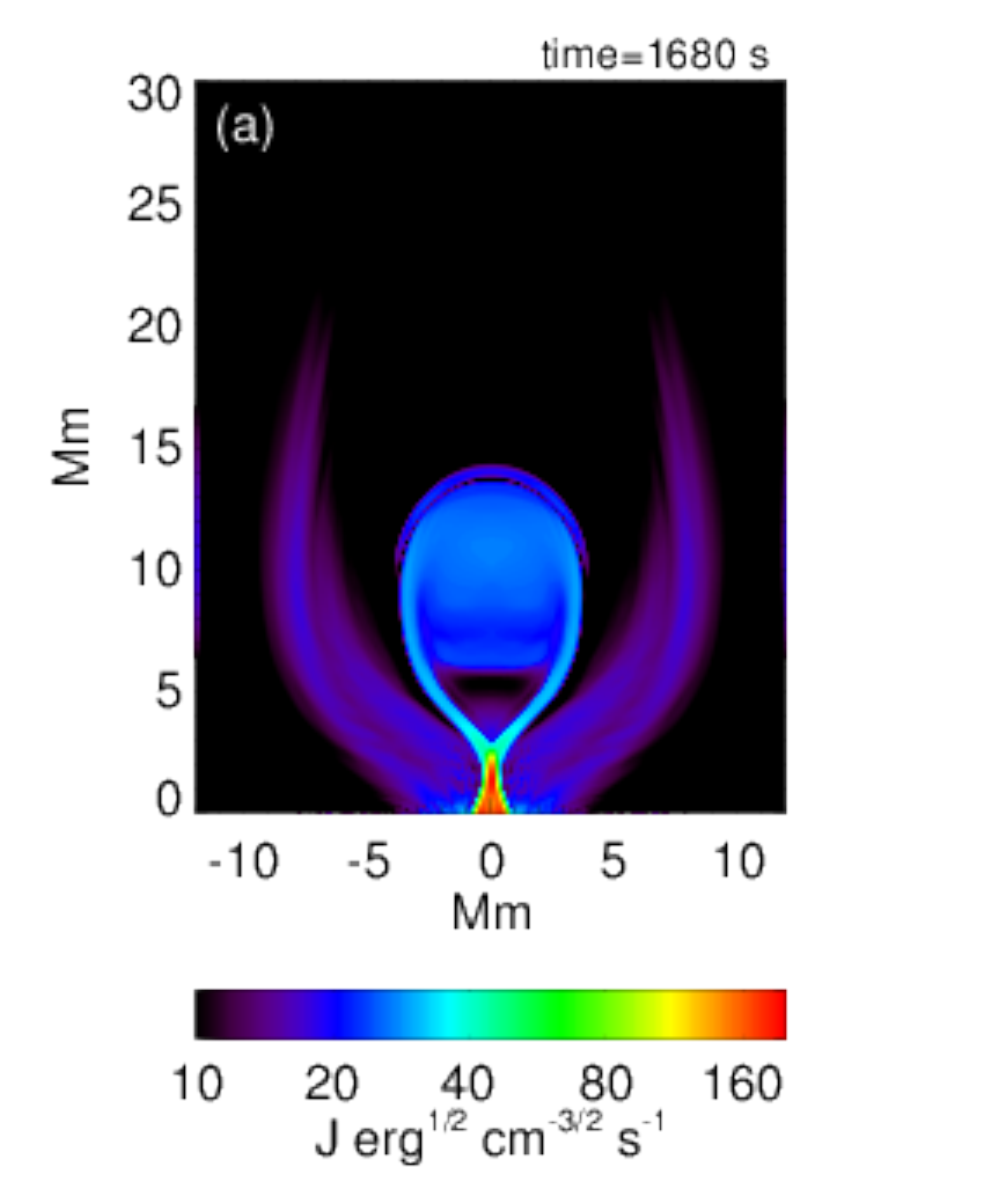}
        \end{center}
      \end{minipage}
      \begin{minipage}{0.5\hsize}
        \begin{center}
          \includegraphics[bb=0 0 283 340,scale=0.85]{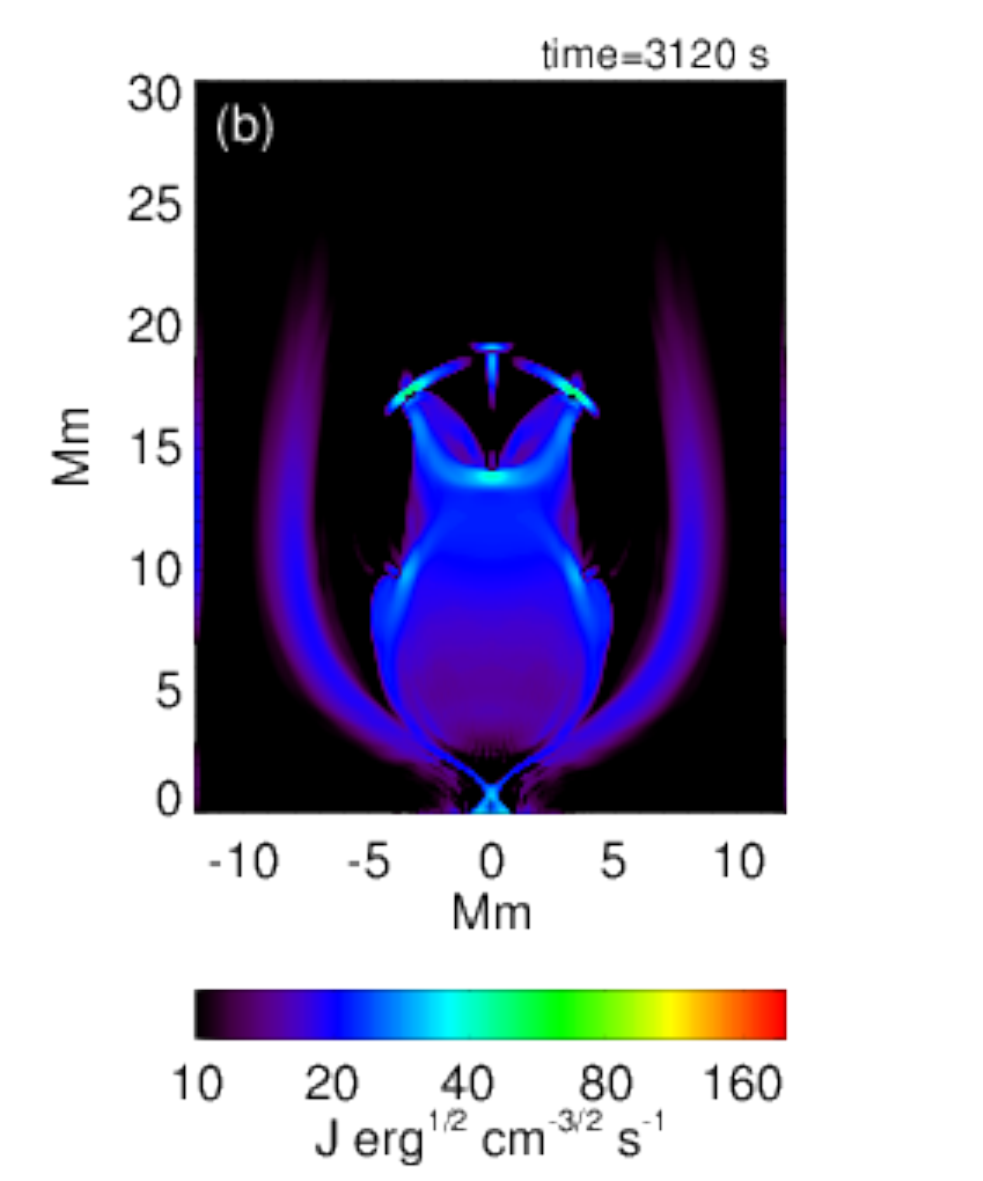}
        \end{center}
      \end{minipage}
    \end{tabular}
    \caption{Current density in $xy$-plane at $z=0$.}
    \label{cur}
  \end{center}
\end{figure}

The reconnection model can affect the initiation of prominence eruption.
In our simulations, the flux rope was elevated by the Petschek type fast reconnection
after anomalous resistivity was swiched on; however, it did not erupt.
Figure \ref{cur} shows snapshots of current density in the $xy$-plane at $z=0$ at
different times. After the onset of reconnection via converging motion,
a localized current sheet typical for the Petschek type reconnection was created
(Fig. \ref{cur} (a)).
The strong current gradually faded away as magnetic flux inside the converging
area was depleted (Fig. \ref{cur} (b)).
The anti-shearing motion also reduced current density.
Eventually, the anomalous resistivity was switched off,
and the elevation of the flux rope was stopped.
2.5-dimensional MHD simulation using uniform resistivity in \citet{Zhao2017ApJ}
succeeded in reproducing prominence eruption by reconnection via converging motion.
In their study, reconnection burst by plasmoid instability in a long current sheet
accelerated the flux rope, leading to eruptions.
Thus the detail of reconnection (or resistivity) model
is one important factor to understand the eruptive mechanism of prominences.

\section{Conclusion}\label{sec:conclusion}
We demonstrated a reconnection--condensation model 
for solar prominence formation 
using three-dimensional MHD simulations including nonlinear anisotropic
thermal conduction and optically thin radiative cooling.
In this model, reconnection and the subsequent topological change 
in the magnetic field cause radiative condensation.
When the length of the reconnected loops exceeds the Field length,
radiative condensation is triggered. The synthesized images
of EUV emissions are consistent with the observational findings,
i.e., in terms of the temporal and spatial shift in the peak
intensities of multiwavelength EUV emissions
from coronal temperatures to prominence temperatures
and the formation of a dark cavity. 
Previous studies have proposed a reconnection scenario
for the formation of a flux rope sustaining a prominence  
\citep{vanBallegooijen1989ApJ,MartensZwaan2001ApJ,Welsch2005ApJ}; 
however, these studies have not been able to explain the origin of cool dense plasmas 
in this reconnection scenario alone.
Our study provides a clear link between reconnection and radiative condensation
and verifies that reconnection leads not only to the flux rope formation 
but also to the generation of cool dense plasmas of prominences 
under the condition associated with the Field length.

\acknowledgements
TK is supported by the Program for Leading Graduate School, FMSP,
MEXT, Japan. This work is supported by JSPS KKAKENHI Grant Number JP16J06780,
JP15H03640, and JP16H03954.
Numerical computations were conducted on a Cray XC30 supercomputer at the
Center for Computational Astrophysics (CfCA) of the National Astronomical
Observatory of Japan.

\end{document}